\documentstyle[12pt]{article}
\newcommand\be{\begin{equation}}
\newcommand\ee{\end{equation}}
\newcommand\bea{\begin{eqnarray}}
\newcommand\eea{\end{eqnarray}}
\newcommand\ket[1]{|#1\rangle}

\newcommand\braket[2]{\langle #1|#2\rangle}
\newcommand{\fatalpha}{{\bf \alpha \kern -0.44em \alpha}}
\newcommand{\fatsigma}{{\bf \sigma \kern -0.54em \sigma}}
\newcommand{\tpchi}{{\bf \chi \kern -0.35em \chi}}
\newcommand{\llambda}{{\bf \lambda \kern -0.45em \lambda}}

\renewcommand{\theequation}{\arabic{equation}}
\renewcommand{\theequation}{\thesection-\arabic{equation}}
\bibliography{plain}
\title{\bf Investigation of continuous-time quantum walk by using Krylov
subspace-Lanczos algorithm   }\vspace{20mm}
\author{ M. A. Jafarizadeh$^{a,b,c}$
 \thanks{E-mail:jafarizadeh@tabrizu.ac.ir}  ,
 S. Salimi$^{a,b}$
 \thanks{E-mail:shsalimi@tabrizu.ac.ir}    ,
 R. Sufiani$^{a,b}$
  \thanks{E-mail:sofiani@tabrizu.ac.ir}
\\ $^a${\small Department of Theoretical Physics and Astrophysics,
Tabriz University, Tabriz 51664, Iran.} \\ $^b${\small Institute
for Studies in Theoretical Physics and Mathematics, Tehran
19395-1795, Iran.} \\ $^c${\small Research Institute for
Fundamental Sciences, Tabriz 51664, Iran.}} \pagebreak

\vspace{20mm}
\begin{document}
\maketitle \vspace{15mm}
\newpage
\begin{abstract}

In  papers\cite{js,jsa}, the amplitudes of continuous-time quantum
walk on graphs possessing quantum decomposition (QD graphs) have
been calculated by a new method  based on  spectral distribution
associated to their adjacency matrix. Here in this paper, it is
shown that the continuous-time quantum walk on any arbitrary graph
can be investigated by spectral distribution method, simply by
using  Krylov subspace-Lanczos algorithm to generate orthonormal
bases of Hilbert space of quantum walk isomorphic to orthogonal
polynomials. Also new type of graphs possessing generalized
quantum decomposition have been introduced, where this is achieved
simply by relaxing some of the constrains imposed on QD graphs and
it is shown that  both in QD and GQD graphs, the unit vectors of
strata are identical with the orthonormal basis produced by
Lanczos algorithm. Moreover, it is shown that probability
amplitude of observing walk at a given vertex is proportional to
its coefficient in the corresponding unit vector of its stratum,
and it can be written in terms of the amplitude of its stratum.
Finally the capability of Lanczos-based algorithm for evaluation
of walk on arbitrary graphs ( GQD or non-QD types), has  been
tested  by calculating the probability amplitudes of quantum walk
on some interesting finite (infinite) graph of GQD type and finite
(infinite) path graph of non-GQD type, where  the asymptotic
behavior of the probability amplitudes at infinite limit of
number of vertices,  are in agreement with those of central limit
theorem of Ref.\cite{nko}.


{\bf PACs Index: 03.65.Ud }
\end{abstract}

\vspace{70mm}
\newpage
\section{Introduction}
Random walks on graphs are the basis of a number of classical
algorithms. Examples include 2-SAT  (satisfiability for certain
types of Boolean formulas), graph connectivity, and finding
satisfying assignments for Boolean formulas.  It is this success
of random walks that motivated the study of their quantum analogs
in order to explore whether they might extend the set of quantum
algorithms.

Recently, the quantum analogue of classical random walks has been
studied in a flurry of works \cite{fg,cfg,abnvw,aakv,mr,k}. The
works of Moore and Russell \cite{mr} and Kempe \cite{k} showed
faster bounds on instantaneous mixing and hitting times for
discrete and continuous quantum walks on a hypercube (compared to
the classical walk).

A study of quantum walks on simple graph is well known in
physics(see \cite{fls}). Recent studies of quantum walks on more
general graphs were described in \cite{fg,cfg,aakv,adz,ccdfgs}.
Some of these works studies the problem in the important context
of algorithmic problems on graphs and  suggests that quantum walks
is a promising algorithmic technique for designing future quantum
algorithms. One approach for investigation of continuous-time
quantum walk on graphs is using the spectral distribution
associated with the adjacency matrix of graphs. Authors in
\cite{js,jsa} have introduced a new method for calculating the
probability amplitudes of quantum walk based on spectral
distribution, where  a canonical relation between the Hilbert
space of stratification corresponding to the graph and a system of
orthogonal polynomials  has been established, which leads to the
notion of quantum decomposition (QD) introduced in \cite{nob,obah}
for the adjacency matrix of graph. Also it is shown in \cite{js}
that by using spectral distribution one can approximate long time
behavior of continuous-time quantum walk on infinite graphs with
finite ones and vice versa. In \cite{js,jsa}, only the particular
graphs of QD type  have been studied .

Here in this work we try to investigate continuous-time quantum
walk on arbitrary graphs by spectral distribution method. To this
aim, first by turning  the graphs into a metric space based on
distance function, we have been able to generalize the
stratification and quantum decomposition  introduced in
\cite{nob}, such that the basis of Hilbert space of quantum walk
consist of superposition of quantum kets of vertices belonging to
the same stratum, but with different coefficients, while the
coefficients are the same in QD case, therefore QD graphs
introduced in \cite{js,nob} are particular kind of graphs
possessing generalized quantum decomposition (GQD).
 Then we show that both in QD and GQD graphs, the unit vectors of strata
are identical with the orthonormal basis produced by Lanczos
algorithm. Also, in the case of GQD graphs we show that
probability amplitude of observing walk at a given vertex is
proportional to its coefficient in corresponding unit vector of
its stratum, and it can be written in terms of the amplitude of
its stratum. For more general graphs, the Lanczos algorithm
transforms the adjacency matrix into a tridiagonal form (quantum
decomposition) iteratively, where we use this fact for studying
non-QD type graphs. Indeed, the Lanczos algorithm gives a
three-term recursion structure to the graph, so the spectral
distribution associated to adjacency matrix can be determined by
Stieltjes transform. In order to see the power of Lanczos-based
algorithm in the investigation of continuous-time quantum walk on
arbitrary graphs ( GQD or non-QD types), we have calculated the
amplitudes of quantum walk on some interesting finite (infinite)
graph of GQD type and finite (infinite) path graph of non-GQD
type.

 The organization of the paper is as follows. In section
2, we review the Krylov subspace methods and Lanczos algorithm. In
Section $3$,  we give a brief outline of some of the main
features of graphs and introduce generalized stratification.
Section $4$ is concerned with the Hilbert space of generalized
stratification. In Section $5$, we review the Stieltjes transform
method for obtaining  spectral distribution $\mu$, and establish
an isometry between orthogonal polynomials and Hilbert space of
generalized stratification. Section $6$ is devoted to the method
for computing amplitudes of continuous-time quantum walk, through
spectral distribution $\mu$ of the adjacency matrix $A$.
 In section $7$  we  calculate the amplitudes of quantum walk on
some interesting finite (infinite) graph of GQD type and finite
(infinite) path graph of non-GQD type. At the end we study  the
asymptotic  behavior the probability amplitudes at infinite limit
of number of vertices, where the results  thus obtained are in
agreement with those of central limit theorem of Ref.\cite{nko}.
 Paper is ended with a brief conclusion together with two appendices.

\section{Krylov subspace-Lanczos algorithm}

In this section we give a brief review of  some of the main
features of Krylov subspace projection methods  and Lanczos
algorithm and more details are  referred to
\cite{bpa,jwi,ltdb,jcrw}.

 Krylov subspace projection methods (KSPM) are
 probably the most
important class of projection methods for linear systems and for
eigenvalue problems. In  KSPM, approximations to the desired
eigenpairs of an $n\times n$ matrix $A$ are extracted from a
$d$-dimensional Krylov subspace
\begin{equation}
 K_d(\ket{\phi_0},A) = span\{\ket{\phi_0},A\ket{\phi_0}, \cdots,A^{d-1}\ket{\phi_0}\},
 \end{equation}
  where $\ket{\phi_0}$
is often a randomly chosen starting vector called reference state
and $d \ll n$. In practice, the retrieval of desired spectral
information is accomplished by constructing an orthonormal basis
$V_d \in R^{n\times d}$ of $K_d(\ket{\phi_0},A)$ and computing
eigenvalues and eigenvectors of the $d$ by $d$ projected matrix
$H_d = {P_{V_d}}^TAP_{V_d}$, where $P_{V_d}$ is projection
operator to $d$-dimensional subspace spanned by the basis $V_d$.

The most popular algorithm for finding an orthonormal basis for
the Krylov subspace, is Lanczos algorithm. The Lanczos algorithm
transforms a Hermitian matrix $A$ into a tridiagonal form
iteratively, i.e., the matrix $A$ will be of tridiagonal form in
the $d$-dimensional projected subspace $H_d$. In fact, the Lanczos
algorithm is deeply rooted in the theory of orthogonal
polynomials, which builds an orthonormal sequence of vectors
$\{\ket{\phi_0},\ket{\phi_1},...,\ket{\phi_{d-1}}\}$ and satisfy
the following three-term recursion relations
\begin{equation}\label{trt}
A\ket{\phi_i}=\beta_{i+1}\ket{\phi_{i+1}}+\alpha_i\ket{\phi_i}+\beta_i\ket{\phi_{i-1}}.
\end{equation}

 The vectors $\ket{\phi_i}, i=0,1,...,d-1$ form an
orthonormal basis for the Krylov subspace $K_d(\ket{\phi_0},A)$.
In these basis, the matrix $A$ is projected to the following
symmetric tridiagonal matrix:
$$L_j=\left(
\begin{array}{ccccc}
 \alpha_0 & \beta_1 & 0 & ... &... \\
      \beta_1 & \alpha_1 & \beta_2 & 0 &... \\
      0 & \beta_2 & \alpha_3 & \beta_3 & ... \\
     ... & ... &...& ... &... \\
     ...& ... &0 & \beta_{d-1} & \alpha_{d-1}\\
\end{array}
\right),$$  where the scalars $\beta_{i+1}$ and $\alpha_i$ are
computed to satisfy two requirements, namely that
$\ket{\phi_{i+1}}$ be orthogonal to $\ket{\phi_i}$ and that
$\|\ket{\phi_{i+1}}\|= 1$.

In fact, the Lanczos algorithm is a modified version of the
classical Gram-Schmidt orthogonalization process. As it can be
seen, at its heart is an efficient three-term recursion relation
which arises because the matrix
$A$ is real and symmetric.\\
If we define the Krylov matrix $K$ such that the columns of $K$
are Krylov basis $\{A^i\phi_0 ; i=0,...,d-1\}$ as:
$$K:=(\ket{\phi_0}, A\ket{\phi_0}, ... , A^{d-1}\ket{\phi_0}),$$
the application of the orthonormalization process to the Krylov
matrix is equivalent to the construction of an upper triangular
matrix $P$ such that the resulting sequence $\Phi=KP$ satisfies
$\Phi^\dag \Phi=1$. We denote by $\ket{\phi_j}$ and $P_j$
respectively the $j$-th column of $\Phi$ and $P$. Then we have
\begin{equation}
\braket{\phi_0}{P_i^{\dagger}(A)P_j(A)|\phi_0}=\braket{KP_i}{KP_j}=\braket{\phi_i}{\phi_j},
\end{equation}
where $P_i=a_0+a_1A+...+a_iA^i$  is a polynomial of
 degree $i$ in indeterminate $A$.

 In the remaining part of this section   we give an algorithmic outline of the Lanczos
 algorithm, where it will be used in calculation of amplitudes of
 continuous-time quantum walk.

\textbf{Lanczos algorithm } \\
Input: Matrix $A\in R^{n\times n}$, starting vector
$\ket{\phi_0}$,
$\|\ket{\phi_0}\|=1$, scalar $d$\\
Output: Orthogonal basis $\{\ket{\phi_0},...,\ket{\phi_{d-1}}\}$
of Krylov subspace $K_d(\ket{\phi_0},A)$
$$
\beta_0=0, \ket{\phi_0}=\ket{\phi}/\|\ket{\phi}\|
$$
 $$ for\;\;\ i=0,1,2,...$$
$$
\ket{\upsilon_i}=A\ket{\phi_i}
$$
$$
\alpha_i=\braket{ \phi_i}{\upsilon_i}
$$
$$
\ket{\upsilon_{i+1}}=\ket{\upsilon_i}-\beta_{i}\ket{\phi_{i-1}}-\alpha_i\ket{\phi_i}
$$
$$
\beta_{i+1}=\|\ket{\upsilon_{i+1}}\|
$$
$$ if$$
$$
\beta_{i+1}\neq 0
$$
$$
\ket{\phi_{i+1}}=\ket{\upsilon_{i+1}}/\beta_{i+1}
$$
$$else$$
$$\ket{\phi_{i+1}}=0.$$

\section{Graphs, adjacency matrix and generalized stratification}
In this section we give a brief outline of some of the main
features of graphs such as  adjacency matrix, distance function
and then by turning  the graphs into a metric space based on
distance function, we have been able to generalize the
stratification introduced in \cite{nob}. A graph is a pair
$\Gamma=(V,E)$, where $V$ is a non-empty set and $E$ is a subset
of $\{(\alpha,\beta);\alpha,\beta\in V,\alpha\neq \beta\}$.
Elements of $V$ and of $E$ are called \emph{vertices} and
\emph{edges}, respectively. Two vertices $\alpha,\beta\in V$ are
called \emph{adjacent} if $(\alpha,\beta)\in E$, and in that case
we write $\alpha\sim \beta$. For a graph $\Gamma=(V,E)$ we define
the adjacency matrix $A$ by

\[
A_{\alpha\beta} = \left\{
\begin{array}{ll}
1 & \mbox{if $ \alpha\sim \beta$}\\
0 & \mbox{otherwise.}
\end{array}
\right.
\]
Obviously, (i) $A$ is symmetric; (ii) an element of $A$ takes a
value in $\{0, 1\}$; (iii) a diagonal element of $A$ vanishes.
Conversely, for a non-empty set $V$, a graph structure is uniquely
determined by such a matrix indexed by $V$.

 The \emph{degree} or
\emph{valency} of a vertex $\alpha\in V$ is defined by
$$
\kappa(\alpha)=|\{\beta\in V; \alpha\sim \beta\}|,
$$
where $\mid.\mid$ denotes the cardinality and $\kappa(\alpha)$ is
finite for all $\alpha\in V$ (local boundedness). A finite
sequence $\alpha_0, \alpha_1, . . . , \alpha_n \in V$ is called a
walk of length $n$ (or of $n$ steps) if $\alpha_{k-1}\sim
\alpha_k$ for all $k=1, 2, . . . , n$. For $\alpha\neq \beta$ let
$\partial(\alpha, \beta)$ be the length of the shortest walk
connecting $\alpha$ and $\beta$. By definition $\partial(\alpha,
\alpha)=0$ for all $\alpha\in V$ and $\partial(\alpha, \beta)=1$
if and only if $\alpha\sim \beta$. Therefore, graphs become
metric space with respect to above defined distance function
$\partial$.

Now, in the remaining part of this section we try to define
generalized stratification based on distance function. To this
aim, similar to association  scheme \cite{bailey} we define a
partition (called distance partition) on $V\times V$, i.e.,
$V\times V=\bigcup_{i}\Gamma_i$  based on distance function
$\partial$, where the subset $\Gamma_i$ are defined by
\begin{equation}
\Gamma_i=\{(\alpha,\beta)\in V\times V
|\partial(\alpha,\beta)=i\}.
\end{equation}
Using above distance partition one can define the set
$\Gamma_i(\alpha)$ ($i$-th neighborhood of vertex $\alpha$) as
\begin{equation}\label{dr}
\Gamma_i(\alpha)=\{\beta\in V | (\alpha, \beta)\in \Gamma_i\}.
\end{equation}
Obviously the class of subsets $\Gamma_i(\alpha)$ defined above
partition $V$ as
\begin{equation}\label{v1}
V=\bigcup_{i}\Gamma_{i}(\alpha),
\end{equation}
(see Fig.1). As we see the graph is stratified into a disjoint
union of strata, hence we call it the generalized stratification
based on distance function with respect to vertex $\alpha$, where
the vertex $\alpha$ is referred to as a reference state(see
Fig.2).

In this stratification for any connected graph $\Gamma$ , we have
\begin{equation}
\Gamma_1(\beta)\subseteq \Gamma_{i-1}(\alpha)\cup
\Gamma_i(\alpha)\cup \Gamma_{i+1}(\alpha),
\end{equation}
 for
each $\beta\in \Gamma_i(\alpha)$.

 Obviously above relations are
similar to those of distance regular graphs \cite{bailey},  where
in the later case the sets $\Gamma_i$ form an association scheme
and the stratification $\Gamma_i(\alpha)$ is independent of
reference state $\alpha$, but in an arbitrary graph the
generalized stratification  depends on  the choice of reference
state.

In order to study continuous-time quantum walk on a given graph
via stratification, we  define in the following section a Hilbert
space which is suitable for Lanczos algorithm.

\section{Hilbert space for the generalized stratification}
Hereafter, we fix a point $o\in V$ as a reference state of the
graph then with each stratum $\Gamma_k(o)$ we associate a unit
vector $\ket{\phi_{k}}$ in $l^2(V)$ called unit vector of $k$-th
stratum.
 The closed subspace of $l^2(V)$ spanned by
$\{\ket{\phi_{k}}\}$ is denoted by $\Lambda(\Gamma)$. In section
$6$, we will deal with continuous-time quantum walk, where
$\Lambda(\Gamma)$ will be referred as walk space denoted by
$V_{walk}$, i.e., the strata $\{\ket{\phi_{k}}\}$ span a closed
subspace, where the quantum walk remains on it forever.

Since $\{\ket{\phi_{k}}\}$ become a complete orthonormal basis of
$\Lambda(\Gamma)$, we often write
\begin{equation}
\Lambda(\Gamma)=\sum_{k}\oplus \textbf{C}\ket{\phi_{k}}.
\end{equation}
Then for each stratum $\Gamma_k(o)$ of generalized stratification,
the unit vector  associated to $k$-th stratum is defined by
\begin{equation}\label{unitv1}
\ket{\phi_{k}}=\frac{1}{\sqrt{\sum_{\alpha}g_{k,\alpha}^2}}\sum_{\alpha\in
\Gamma_{k}(o)}g_{k,\alpha}\ket{k,\alpha},
\end{equation}
where, $\ket{k,\alpha}$ denotes the eigenket of $\alpha$-th vertex
at the stratum $k$ and integers $g_{k,\alpha}\geq 1$ for each
$\alpha\in \Gamma_{k}(o)$. We refer to a  graph as GQD graph if
the coefficients $g_{k,\alpha}$  satisfy  conditions appearing in
appendix $A$ (the conditions  (\ref{A1}) through (\ref{A3})).

By choosing  $g_{k,\alpha}=1$ for each $\alpha\in \Gamma_{k}(o)$,
 Eq.(\ref{unitv1}) reduces to
\begin{equation}\label{unitv}
\ket{\phi_{k}}=\frac{1}{\sqrt{|\Gamma_{k}(o)}|}\sum_{\alpha\in
\Gamma_{k}(o)}\ket{k,\alpha},
\end{equation}
where, $\ket{\phi_{k}}$, $k=0,1,2,...$ correspond to unit vectors
 of QD graphs of Ref.\cite{js}.

In the following we  show that, for the QD type graphs the unit
vectors of strata given in Eq.(\ref{unitv}), are  the same as the
orthonormal basis produced via Lanczos algorithm (this is true for
GQD graphs too, where its proof is referred to appendix $A$.). To
do so, let us consider the action of adjacency matrix $A$ over
$\ket{\phi_{k}}$ as
$$
A\ket{\phi_{k}}=\frac{1}{\sqrt{|\Gamma_{k}(o)|}}\sum_{\alpha\in
\Gamma_{k}(o)}A\ket{k,\alpha}
$$$$
=\frac{1}{\sqrt{|\Gamma_{k}(o)|}}\sum_{\alpha\in
\Gamma_{k}(o)}\sum_{\nu\in \Gamma_{k+1}(o),\nu\sim
\alpha}\ket{k+1,\nu}+\frac{1}{\sqrt{|\Gamma_{k}(o)|}}\sum_{\alpha\in
\Gamma_{k}(o)}\sum_{\nu\in \Gamma_{k}(o),\nu\sim
\alpha}\ket{k,\nu}
$$
$$
+\frac{1}{\sqrt{|\Gamma_{k}(o)|}}\sum_{\alpha\in
\Gamma_{k}(o)}\sum_{\nu\in \Gamma_{k-1}(o),\nu\sim
\alpha}\ket{k-1,\nu}=\sqrt{\frac{|\Gamma_{k+1}(o)|}{|\Gamma_{k}(o)|}}
\frac{1}{\sqrt{|\Gamma_{k+1}(o)|}}\sum_{\nu\in
\Gamma_{k+1}(o)}\lambda_{k+1}(\nu)\ket{k+1,\nu}+
$$
\begin{equation}\label{d1}
 \frac{1}{\sqrt{|\Gamma_{k}(o)|}}\sum_{\nu\in
\Gamma_{k}(o)}\alpha_{k}(\nu)\ket{k,\nu}+\sqrt{\frac{|\Gamma_{k-1}(o)|}{|\Gamma_{k}(o)|}}
\frac{1}{\sqrt{|\Gamma_{k-1}(o)|}}\sum_{\nu\in
\Gamma_{k-1}(o)}\frac{\lambda_{k}(\nu)|\Gamma_{k}(o)|}{|\Gamma_{k-1}(o)|}\ket{k-1,\nu}.
\end{equation}

By defining
$\beta_{k}=\frac{|\Gamma_{k}|^{1/2}}{|\Gamma_{k-1}|^{1/2}}\lambda_{k}(\nu)$,
$\lambda_{k}(\nu)=|\{\alpha\in \Gamma_{k-1}(o);\alpha\sim\nu \}|$
 and
$\alpha_{k}=|\{\nu\in \Gamma_k;\nu\sim \alpha\}|$ for $\alpha,
\nu\in \Gamma_k(o)$, the three-term recursion relations (\ref{d1})
reduce to those given in  (\ref{trt}).

Therefore, the adjacency matrix takes a tridiagonal form in the
basis $\ket{\phi_k}$ (orthonormal  basis associated with strata),
consequently these basis are identical with the orthonormal basis
produced by Lanczos algorithm.
\section{Spectral distribution of the adjacency matrix $A$}
It is well known that, for any pair $(A,\ket{\phi_0})$ of a matrix
$A$ and a vector $\ket{\phi_0}$, it can be assigned a measure
$\mu$ as follows
\begin{equation}\label{sp1}
\mu(x)=\braket{ \phi_0}{E(x)|\phi_0},
\end{equation}
 where
$E(x)=\sum_i|u_i\rangle\langle u_i|$ is the operator of projection
onto the eigenspace of $A$ corresponding to eigenvalue $x$, i.e.,
\begin{equation}
A=\int x E(x)dx.
\end{equation}
It is easy to see that, for any polynomial $P(A)$ we have
\begin{equation}\label{sp2}
P(A)=\int P(x)E(x)dx,
\end{equation}
where for discrete spectrum the above integrals are replaced by
 summation.

Actually the spectral analysis of operators  is an important issue
in quantum mechanics, operator theory and mathematical physics
\cite{simon, Hislop}. As an example $\mu(dx)=|\psi(x)|^2dx$
($\mu(dp)=|\widetilde{\psi}(p)|^2dp$) is a spectral distribution
which is  assigned to  the position (momentum) operator
$\hat{X}(\hat{P})$. Moreover, in general quasi-distributions are
the assigned spectral distributions of two hermitian non-commuting
operators with a prescribed ordering.
 For
example the Wigner distribution in phase space is the assigned
spectral distribution for two non-commuting operators $\hat{X}$
(shift operator) and $\hat{P}$ (momentum operator) with
Wyle-ordering among them \cite{Kim,Hai}.

Here in this paper we are concerned with spectral distribution of
adjacency matrices of graphs, since the spectrum of a given graph
can be determined by spectral distribution of its adjacency
matrix $A$.

Therefore, using the relations (\ref{sp1}) and (\ref{sp2}), the
expectation value of powers of adjacency matrix $A$ over starting
site $\ket{\phi_0}$ can be written as
\begin{equation}\label{v2}
\braket{\phi_{0}}{A^m|\phi_0}=\int_{R}x^m\mu(dx), \;\;\;\;\
m=0,1,2,....
\end{equation}
The existence of a spectral distribution satisfying (\ref{v2}) is
a consequence of Hamburger's theorem, see e.g., Shohat and
Tamarkin [\cite{st}, Theorem 1.2].

Obviously relation (\ref{v2})  implies an isomorphism from the
Hilbert space of generalized stratification onto the closed linear
span of the orthogonal polynomials with respect to the measure
$\mu$. Since, from the orthogonality of vectors $\ket{\phi_j}$ (
Hilbert space of generalized stratification) produced from Lanczos
algorithm process we have,
$$\delta_{ij}=\braket{\phi_i}{\phi_j}=\braket{\phi_0}{P_i^{\dagger}(A)P_j(A)|\phi_0}$$
\begin{equation}
=\int P_i^*(x)P_j(x)\mu(x)dx=(P_i,P_j)_{\mu}.
\end{equation}
 Conversely if $P_0,...,P_{n-1}$ is the system of orthonormal
polynomials for the measure $\mu$ then the vectors
\begin{equation}\label{poly}
\ket{\phi_j}= P_j(A)\ket{\phi_0},
\end{equation}
will coincide with the sequence of orthonormal vectors produced by
the Lanczos algorithm applied to $(A,\ket{\phi_0})$.

Now, substituting (\ref{poly}) in (\ref{trt}), we get three term
recursion relations between polynomials $P_j(A)$, which leads to
 the following  three term recursion between polynomials $P_j(x)$
\begin{equation}
\beta_{k+1}P_{k+1}(x)=(x-\alpha_k)P_{k}(x)-\beta_kP_{k-1}(x)
\end{equation}
for $k=0,...,n-1$.\\
Multiplying by $\beta_1...\beta_k$ we obtain
\begin{equation}
\beta_1...\beta_{k+1}P_{k+1}(x)=(x-\alpha_k)\beta_1...\beta_kP_{k}(x)-\beta_k^2.\beta_1...\beta_{k-1}P_{k-1}(x).
\end{equation}
By rescaling $P_k$ as $P'_k=\beta_1...\beta_kP_k$, the spectral
distribution $\mu$ under question is characterized by the property
of orthonormal polynomials $\{P'_n\}$ defined recurrently by
$$ P'_0(x)=1, \;\;\;\;\;\
P'_1(x)=x,$$
\begin{equation}\label{op}
xP'_k(x)=P'_{k+1}(x)+\alpha_{k}P'_k(x)+\beta_k^2P'_{k-1}(x),
\end{equation}
for $k\geq 1$.

 If such a spectral distribution is unique, the
spectral distribution $\mu$ is determined by the identity:
\begin{equation}\label{v3}
G_{\mu}(z)=\int_{R}\frac{\mu(dx)}{z-x}=\frac{1}{z-\alpha_0-\frac{\beta_1^2}{z-\alpha_1-\frac{\beta_2^2}
{z-\alpha_2-\frac{\beta_3^2}{z-\alpha_3-\cdots}}}}=\frac{Q_{n-1}^{(1)}(z)}{P'_{n}(z)}=\sum_{l=1}^{n}
\frac{A_l}{z-x_l},
\end{equation}
where $G_{\mu}(z)$ is called the Stieltjes transform of spectral
distribution $\mu$ and
 polynomials $\{Q_{k}^{(1)}\}$ are defined recurrently as
        $$Q_{0}^{(1)}(x)=1, \;\;\;\;\;\
    Q_{1}^{(1)}(x)=x-\alpha_1$$,
    \begin{equation}\label{oq}
    xQ_{k}^{(1)}(x)=Q_{k+1}^{(1)}(x)+\alpha_{k+1}Q_{k}^{(1)}(x)+\beta_{k+1}^2Q_{k-1}^{(1)}(x),
   \end{equation}
for $k\geq 1$. The coefficients $A_l$ appearing in (\ref{v3}) are
the same Guass quadrature constants which are calculated as
\begin{equation}\label{Gauss}
A_l=\lim_{z\rightarrow x_l}(z-x_l)G_{\mu}(z),
\end{equation}
where, $x_l$ are the roots  of polynomial $P'_{n}(x)$.

Now let $G_{\mu}(z)$ is known, then the spectral distribution
$\mu$ can be recovered from $G_{\mu}(z)$ by means of the Stieltjes
inversion formula as
\begin{equation}\label{m1}
\mu(y)-\mu(x)=-\frac{1}{\pi}\lim_{v\longrightarrow
0^+}\int_{x}^{y}Im\{G_{\mu}(u+iv)\}du.
\end{equation}
Substituting the right hand side of (\ref{v3}) in (\ref{m1}), the
spectral distribution can be determined in terms of $x_l,
l=1,2,...$ and  Guass quadrature constants $A_l, l=1,2,... $ as
\begin{equation}\label{m}
\mu=\sum_l A_l\delta(x-x_l)
\end{equation}
( for more details see Ref. \cite{obah,st,tsc,obh}).

Finally, using the relation (\ref{poly}) and the recursion
relations (\ref{op}) of polynomial $P'_k(x)$, the other matrix
elements $\label{cw1} \braket{\phi_{k}}{A^m\mid \phi_0}$ can be
calculated as
\begin{equation}\label{cw1}
\braket{\phi_{k}}{A^m\mid \phi_0}=\frac{1}{\beta_1\beta_2\cdots
\beta_k }\int_{R}x^{m}P'_{k}(x)\mu(dx),  \;\;\;\;\ m=0,1,2,....
\end{equation}
\section{Investigation of Continuous-time quantum walk on an arbitrary graph via
spectral distribution  of its adjacency matrix}

Our main goal in this paper is the evaluation of probability
amplitudes  for continuous-time quantum walk by using
Eq.(\ref{cw1}), such that we have
\begin{equation} \label{v4}
q_{k}(t)=\braket{\phi_{k}}{e^{-iAt}\mid
\phi_0}=\frac{1}{\beta_1\beta_2\cdots \beta_k
}\int_{R}e^{-ixt}P'_{k}(x)\mu(dx),
\end{equation}
where $q_{k}(t)$ is the amplitude of observing the walk at stratum
$k$ at time $t$. The conservation of probability $\sum_{k=0}{\mid
q_{k}(t)\mid}^2=1$ follows immediately from Eq.(\ref{v4}), simply
by using the completeness relation of orthogonal polynomials
$P'_k(x)$.

Investigation of continuous-time quantum walk via spectral
distribution method, pave the way to approximate infinite graphs
with finite ones and vice versa, simply via Gauss quadrature
formula, where in cases of infinite graphs, one can study
asymptotic behavior of walk at large enough times by using the
method of stationary phase approximation (for more details see
\cite{js}).

 One should  note
that, the spectral distribution is Fourier transform of the
amplitude of observing the walk at starting site at time $t$,
i.e.,
\begin{equation}
q_0(t)=\int e^{-ixt}\mu(x)dx \;\;\  \longmapsto \;\;\
\mu(x)=\frac{1}{2\pi}\int e^{ixt}q_0(t)dt.
\end{equation}
Above relations imply that
\begin{equation}
q_k(t)=\frac{1}{\beta_1\beta_2\cdots \beta_k }\int
P'_k(x)e^{-ixt}\mu(x)dx=\frac{1}{2\pi\beta_1\beta_2\cdots \beta_k
}\int P'_k(x)q_0(t')e^{-ix(t-t')}dt'dx,
\end{equation}
therefore, the amplitudes $q_k(t)$ can be written  in terms of
the amplitude $q_0(t)$.

 Obviously for finite graphs, the
formula (\ref{v4}) yields
\begin{equation}\label{fin}
q_{k}(t)=\frac{1}{\beta_1\beta_2\cdots \beta_k
}\sum_{l}A_le^{-ix_lt}P'_{k}(x_l),
\end{equation}
where by straightforward  calculation one can  evaluate  the
average probability for the finite graphs as
\begin{equation}
P(k)=\lim_{T\rightarrow \infty}\frac{1}{T}\int_{0}^{T}\mid
q_{k}(t)\mid ^2dt=\frac{1}{\beta_1\beta_2\cdots \beta_k
}\sum_{l}A_l^2{P'}^2_{k}(x_l).
\end{equation}
In appendix I of Ref.\cite{js} it is proved that for QD graphs the
amplitudes on the vertices belonging to the one stratum is the
same, hence the probability of observing the walk at a site
belonging to stratum $k$ is equal to
$\frac{|q_{k}(t)|^2}{|\Gamma_k(o)|}$. Unfortunately for non-QD
graphs the lemma appearing in  appendix I of Ref.\cite{js} is not
true any more, consequently the probability amplitudes of
observing walk at sites can not be obtained  from those of strata
in a simple way and reader can follow  the details of calculation
of amplitudes in appendix $B$.
\section{Examples}

\subsection{Generalized QD graphs}
Here in this subsection we give  examples of GQD type graphs.
These graphs look like  kite and they are embedded in $Z^k,
k=2,3,...$ lattices and defined as follows: Let K(k,n) be an
k-dimensional lattice graph with $n+1$ generalized strata, which
consists of vertices $\underbrace{(0,0,...,0)}_k,(l,0,...,0),
(0,l,0,...,0),..., (0,0,...0,l)$ and $\underbrace{(l,l,...,l)}_k$
only for odd values of $l$, where $l=0,1,...,n$. The vertex
$(0,0,...,0)$ is connected to vertices
$(1,0,...,0),(0,1,...,0),...,(0,0,...,1)$ , the vertex $(0, ...,
0, \underbrace{l}_i, 0,..., 0)$ is connected to vertices $(0,
..., 0, \underbrace{l-1}_i, 0,..., 0)$ and $(0,
 ..., 0, \underbrace{l+1}_i, 0, ..., 0)$ for each $i=1,...,k$, but for
odd values of $l$, there is an extra connection between $(0,...,0,
\underbrace{l}_i, 0,..., 0)$ and $(l, l, ..., l)$ (see Fig.3 for
$k=2$, $n=6$).

Now, we define unit vectors of generalized strata  in such a way
that, they coincide with the orthonormal basis produced by lanczos
algorithm (see appendix $A$)
$$
\ket{\phi_0}=\ket{\underbrace{0, 0, ..., 0}_k}
$$
$$
\ket{\phi_1}=\frac{1}{\sqrt{k}}\sum_{perm.}\ket{1, 0, ..., 0}
$$
$$
\ket{\phi_2}=\frac{1}{\sqrt{k(k+1)}}(\sum_{perm.}\ket{2,
0,...,0}+k\ket{1, 1, ..., 1})
$$
$$
\vdots
$$
$$
\ket{\phi_{2l-1}}=\frac{1}{\sqrt{k}}\sum_{perm.}\ket{2l-1,
0,...,0}
$$
\begin{equation}
\ket{\phi_{2l}}=\frac{1}{\sqrt{k(k+1)}}(\sum_{perm.}\ket{2l,
0,..., 0}+k\ket{2l-1, 2l-1,..., 2l-1}),
\end{equation}
where, the summations are taken over all possible permutations.
Using the  relations (\ref{A1})-(\ref{A3}), one can show that the
coefficients $\beta_i$ and $\alpha_i$ are
\begin{equation}\label{ab}
\beta_1^2=k, \;\;\;\;\ \beta_2^2=\beta_3^2=....=k+1 \;\;\
\mbox{and} \;\;\ \alpha_i=0 , \;\ i=1,2,....
\end{equation}
Now, one can study quantum continuous time walk on these graphs
for finite values of $n$ simply by following the general
prescriptions, but here we restrict ourselves to infinite $n$.
Substituting coefficients $\beta_i$ and $\alpha_i$ in (\ref{v3})
, the Stieltjes transform $G_{\mu}(z)$ of spectral distribution
$\mu$ takes the following form
\begin{equation}
G_{\mu}(z)=\frac{1}{z-\frac{k}{z-\frac{k+1}
{z-\frac{k+1}{z-\cdots}}}}.
\end{equation}
In order to evaluate above infinite continued fraction, we need
first to evaluate the following infinite continued fraction
defined as
\begin{equation}\label{st.10}
\tilde{G}(z)=\frac{1}{z-\frac{k+1}{z-\frac{k+1}
{z-\frac{k+1}{z-\cdots}}}}= \frac{1}{z-(k+1)\tilde{G}(z)},
\end{equation}
where by solving above equation, we get
\begin{equation}\label{st.1}
\tilde{G}(z)=\frac{z-\sqrt{z^2-4(k+1)}}{2(k+1)}.
\end{equation}
Inserting (\ref{st.1}) in (\ref{st.10}), we get
\begin{equation}\label{st.3}
G_{\mu}(z)=\frac{1}{z-kG'_{\mu}(z)},
\end{equation}
 then substituting (\ref{st.1}) in (\ref{st.3}), we obtain
the following expression for Stieltjes transform of $\mu$
\begin{equation}\label{st.2}
G_{\mu}(z)=\frac{(k+2)z-k\sqrt{z^2-4(k+1)}}{2(k^2+z^2)},
\end{equation}
 finally by applying Stieltjes
inversion formula, we get the absolutely continuous part of
spectral distribution $\mu$ as follows
\begin{equation}
\mu(x)=\frac{k}{2\pi}\frac{\sqrt{4(k+1)-x^2}}{k^2+x^2}, \;\;\;\;\
|x|\leq 2\sqrt{k+1} .
\end{equation}
Now, we study the  probability amplitudes  of walk at time $t$ in
the limit of large $k$ i.e.,
$$
q_l(t)=\lim_{k\rightarrow\infty}\braket{\phi_l}{e^{\frac{-iAt}{\sqrt{k}}}\mid\phi_0}=
\lim_{k\rightarrow\infty}\frac{1}{\sqrt{k(k+1)^l}}
\int_{-2\sqrt{k+1}}^{2\sqrt{k+1}}e^{\frac{-ixt}{\sqrt{k}}}P'_l(x)
\frac{k}{2\pi}\frac{\sqrt{4(k+1)-x^2}}{k^2+x^2}dx
$$
\begin{equation}\label{cent}
=\lim_{k\rightarrow\infty}\frac{1}{2\pi\sqrt{k(k+1)^l}}
\int_{-2\sqrt{(k+1)/k}}^{2\sqrt{(k+1)/k}}e^{-ixt}P'_l(\sqrt{k}x)\frac{\sqrt{4(k+1)/k-x^2}}{1+x^2/k}dx
\end{equation}
$$
=\frac{1}{2\pi}\int_{-2}^{2}e^{-ixt}P'_{l,\infty}(x)\sqrt{4-x^2}dx
=\frac{2}{\pi}\int_{-1}^{1}e^{-i2xt}P'_{l,\infty}(2x)\sqrt{1-x^2}dx,
$$

where the  polynomial $P'_{l,\infty}(x)$  is defined by
\begin{equation}\label{Eq3}
P'_{l,\infty}(x)=\lim_{k\rightarrow\infty}\frac{1}{\sqrt{k(k+1)^{l-1}}}P'_l(\sqrt{k}x).
\end{equation}
Now, substituting $\beta_i$ and $\alpha_i$ from (\ref{ab}) in
three-term recursion relations (\ref{op}), we obtain the following
relations for polynomials $P'_l(x)$
$$
P'_{0}(\sqrt{k}x)=1,
$$
$$
P'_{1}(\sqrt{k}x)=\sqrt{k}x,
$$
$$
P'_{2}(\sqrt{k}x)=kx^2-k,
$$
\begin{equation}\label{Eq4}
\sqrt{k}xP'_{l}(\sqrt{k}x)=P'_{l+1}(\sqrt{k}x)+(k+1)P'_{l-1}(\sqrt{k}x),
\;\;\ l=3,4,...\;\ .
\end{equation}
Then  dividing left and right hand sides of the recursion
relations in (\ref{Eq4}) by $\sqrt{k(k+1)^{l-1}}$ and taking the
limit at $k\rightarrow\infty$, one can obtain the following
recursion relations for $P'_{l,\infty}(x)$
$$
P'_{0,\infty}(x)=1,
$$
$$
P'_{1,\infty}(x)=\lim_{k\rightarrow\infty}\frac{\sqrt{k}x}{\sqrt{k}}=x,
$$
$$
P'_{2,\infty}(x)=\lim_{k\rightarrow\infty}\frac{kx^2-k}{\sqrt{k(k+1)}}=x^2-1,
$$
\begin{equation}\label{Eq5}
xP'_{l,\infty}(x)=P'_{l+1,\infty}(x)+P'_{l-1,\infty}(x),
 \;\;\l=3,4,...\;\ .
\end{equation}
By comparing the recursion relations (\ref{Eq5}) of
$P'_{l,\infty}(x)$ with those of Tchebichef polynomials of second
kind, we conclude that
\begin{equation}
P'_{l,\infty}(x)=U_l(x/2),
\end{equation}
 where, $U_l(x)$'s are Tchebichef polynomials of second
kind. Therefore the probability amplitudes in Eq.(\ref{cent}) can
be rewritten as
\begin{equation}\label{Eq1}
q_l(t)=\frac{2}{\pi}\int_{-1}^1
e^{-2ixt}U_l(x)\sqrt{1-x^2}dx=\frac{2}{\pi}\int_{-1}^1
e^{-2ixt}\sin((l+1)\cos^{-1}x)dx.
\end{equation}
Now, by doing the change the variable $x=\cos\theta$, the
integral(\ref{Eq1}) can be written as
\begin{equation}\label{Eq2}
q_l(t)=\frac{2}{\pi}\int_{0}^{\pi}
e^{-2it\cos{\theta}}\sin((l+1)\theta)\sin\theta d\theta.
\end{equation}
Then, using the following integral representation of Bessel
polynomials
\begin{equation}
J_l(x)=\frac{i^{-l}}{\pi}\int_0^{\pi}e^{-ix\cos\theta}\cos\theta
d\theta,
\end{equation}
the integral in (\ref{Eq2}) can be written as
\begin{equation}
q_l(t)=i^l(J_l(2t)+J_{l+2}(2t)).
\end{equation}
Now, from the recursion relations for Bessel polynomials, i.e.,
\begin{equation}
J_{l+1}(x)=\frac{2l}{x}J_l(x)-J_{l-1}(x),
\end{equation}
we obtain the following expression for the probability amplitudes
of  walk in the limit of large $k$
\begin{equation}
q_l(t)=(l+1)i^l\frac{J_{l+1}(2t)}{t},
\end{equation}
where the results are in agreement with the corresponding quantum
central limit theorem of Ref.\cite{nko}.

\subsection{Non-GQD type graphs} In this subsection we study an example of non-GQD type
graphs, those graphs that do not possess three term recursion
property. In order to obtain spectral distribution of adjacency
matrix of a give non-GQD  graph, we need to find the  basis in
which the adjacency matrix has tridiagonal form. To this aim  we
have to choose starting site of walk as a reference state and then
apply Lanczos algorithm to its adjacency matrix.Then by using
spectral distribution, we will be able to calculate the
amplitudes of walk as will be explained in the following example.
  \subsubsection{Walk on finite path graph with second vertex as the  starting
site of the walk} Finite path graph $\textsf{P}_n=\{1,2,... \}$
 is a $n$- vertex graph with $n-1$ edges  all on a single
open path \cite{js}. For this graph, the stratification depends on
the choice of  starting site of walk. If we choose the  second
vertex as starting site of the walk, as it is shown in Fig.2, the
graph does not satisfy a three term recursion relations, i.e., the
adjacency matrix has not tridiagonal form.

Therefore, in order to find the  basis in which the adjacency
matrix has tridiagonal form,  we have to  apply Lanczos algorithm
to the adjacency matrix $A$ of the graph $\textsf{P}_n$, where
starting site $|\phi_0\rangle=\ket{1}$ is chosen as a reference
state.    Also, the Lanczos algorithm provides the coefficients
$\alpha$ and $\beta$ from which the Stieltjes transform
$G_\mu(z)$ of $\mu$, Eq.(\ref{v3}) can be calculated.

Hence, following the prescription of Lanczos algorithm given in
section $2$, we get the following results for $\textsf{P}_n$,
which are different for  even and odd values of $n$.
  \\
\textbf{A.} $n=2k$\\
$\alpha_i=0,\;\;\;\;\ i=0,1,...,2k-1,$\\
$\beta_{2i}=\sqrt{\frac{i}{i+1}}, \;\;\;\ \\
\beta_{2i-1}=\sqrt{\frac{i+1}{i}},\;\;\;\;\ i=1,...,k-1,$\\
$\beta_{2k-1}=\frac{1}{\sqrt{k}}.$\\
\textbf{B.} $n=2k+1$\\
$\alpha_i=0,\;\;\;\;\ i=0,1,...,2k-1,$\\
$\beta_{2i}=\sqrt{\frac{i}{i+1}},\;\;\;\ i=1,..., k-1 \\
\beta_{2i-1}=\sqrt{\frac{i+1}{i}},\;\;\;\;\ i=1,...,k,$
respectively.

Substituting the coefficients $\alpha_i$ and $\beta_i$ in
(\ref{op}) and (\ref{oq}), and using (\ref{v3}), we get the
following  closed form of the Stieltjes transform of $\mu$
\begin{equation}
G_\mu(z)=\frac{zU_{n-2}(z/2)}{U_n(z/2)}
\end{equation}
 where, $U_n$'s are Tchebichef polynomials of second kind.
 Therefore, the roots $x_l$ appearing in (\ref{v3}) are roots of Tchebichef
polynomials of second kind, i.e., $x_l=2\cos(\frac{l\pi}{n+1})$.
 Also,  using (\ref{Gauss}) we get the following expression for the coefficients $A_l$
\begin{equation}
A_l=\frac{2}{n+1}\sin^2(\frac{2l\pi}{n+1}).
\end{equation}
Thus,  spectral distribution is given by
\begin{equation}
\mu=\frac{2}{n+1}\sum_{l=1}^{n}\sin^2(\frac{2l\pi}{n+1})\delta(x-2\cos(\frac{l\pi}{n+1})).
\end{equation}
Then the probability amplitude  of the walk at starting site at
time $t$ is
\begin{equation}
q_0(t)
=\frac{1}{n+1}\sum_{l=1}^{n}\sin^2(\frac{2l\pi}{n+1})e^{-2it\cos{l\pi/(n+1)}},
\end{equation}
again one can calculate the other amplitudes by using
Eq.(\ref{fin}).

 It should be noticed  that, for odd $n$ the Lanczos algorithm
produces $n-1$ orthonormal basis, therefore for calculating the
amplitudes on vertices we need to construct an extra vector
orthogonal to the walk space $V_{walk}$.

Finally, in the limit of large $n$,  the  continuous part of
spectral distribution $\mu(x)$ is obtained as follows
$$
 \mu(x)=\frac{2}{\pi}\int_{0}^{\pi}
dy\sin^2(2y)\delta(x-2\cos(y))
$$
 $$=\frac{2}{\pi}\int_{0}^{\pi}
dy\frac{\sin^2(2y)\delta(y-\arccos(x/2))}{2\sin(y)}
$$
$$
=\frac{4}{\pi}\int_{0}^{\pi}
dy\sin(y)\cos^2(y)\delta(y-\arccos(x/2))
$$
\begin{equation}
=\frac{1}{2\pi}x^2\sqrt{4-x^2},  \;\;\;\;\;\;\ -2\leq x\leq 2,
\end{equation}
therefore, the probability amplitude of the walk at starting site
at time $t$ is
\begin{equation}
q_0(t)=\frac{1}{2\pi}\int_{-2}^{2}e^{-ixt}x^2\sqrt{4-x^2}dx=\frac{4J_1(2t)}{t}-\frac{6J_2(2t)}{t^2},
\end{equation}
where  above result is obtained by  making  the change of variable
$x=\cos\theta$, and using the  integral representation of Bessel
polynomials given in (7-47).  Similarly, other amplitudes of walk
can be calculated by using Eq.(\ref{v4}).

\section{Conclusion}
 By turning  the graphs into a metric space based on
distance function, we have been able to generalize the
stratification and quantum decomposition  introduced in
\cite{nob}. Then the continuous-time quantum walk on arbitrary
graphs are investigated by spectral distribution method based on
Krylov subspace-Lanczos algorithm. We  have showed that both in QD
and GQD graphs, the unit vectors of strata are identical with the
orthonormal basis produced by Lanczos algorithm.  For more
general graphs, we have used the Lanczos algorithm to get a basis
in which the adjacency matrix has tridiagonal form, where it is
necessary for determination of spectral distribution of adjacency
matrix by using inverse Stieltjes transform. We believe that the
introduced algorithm is a powerful and general tool to
investigate the continuous-time quantum walk on any arbitrary
graph.

\vspace{1cm} \setcounter{section}{0}
 \setcounter{equation}{0}
 \renewcommand{\theequation}{A-\roman{equation}}
  {\Large{Appendix A}}

  In this appendix we  show that in the case of GQD graphs the unit
vectors of strata (i.e., Eq.(\ref{unitv1})), are the same as the
orthonormal basis produced via Lanczos algorithm. To do so, let us
consider the action of adjacency matrix $A$ over $\ket{\phi_{k}}$
as
$$
A\ket{\phi_{k}}=\frac{1}{\sqrt{\sum_{\alpha}g_{k,\alpha}^2}}\sum_{\alpha\in
\Gamma_{k}(o)}g_{k,\alpha}A\ket{k,\alpha}
=\frac{1}{\sqrt{\sum_{\alpha}g_{k,\alpha}^2}}\sum_{\alpha\in
\Gamma_{k}(o)}g_{k,\alpha}\sum_{\nu\in
\Gamma_{k+1}(o),{\alpha\sim\nu}}\ket{k+1,\nu}
$$
\begin{equation}\label{tree1}
+\frac{1}{\sqrt{\sum_{\alpha}g_{k,\alpha}^2}}\sum_{\alpha\in
\Gamma_{k}(o)}g_{k,\alpha}\sum_{\nu\in
\Gamma_{k}(o),{\alpha\sim\nu}}\ket{k,\nu}
+\frac{1}{\sqrt{\sum_{\alpha}g_{k,\alpha}^2}}\sum_{\alpha\in
\Gamma_{k}(o)}g_{k,\alpha}\sum_{\nu\in
\Gamma_{k-1}(o),{\alpha\sim\nu}}\ket{k-1,\nu},
\end{equation}
now in order to have a GQD graph the coefficients $g_{k,\alpha}$
should satisfy the following conditions
\begin{equation}\label{A1}
\sum_{\alpha\in\Gamma_{k}(o)}g_{k,\alpha}=\gamma_{k+1}g_{k+1,\nu},
\end{equation}
for all $\nu\in\Gamma_{k+1}(o)$ and $\alpha\sim\nu$,
\begin{equation}\label{A2}
\sum_{\alpha\in\Gamma_{k}(o)}g_{k,\alpha}=\eta_k g_{k,\nu},
\end{equation}
for all $\nu\in\Gamma_{k}(o)$ and $\alpha\sim\nu$,
\begin{equation}\label{A3}
\sum_{\alpha\in\Gamma_{k}(o)}g_{k,\alpha}=
\gamma_{k}(\frac{\sum_{\alpha\in\Gamma_{k}(o)}g_{k,\alpha}^2}
{\sum_{\xi\in\Gamma_{k-1}(o)}g_{k-1,\xi}^2})g_{k-1,\nu},
\end{equation}
for all $\nu\in\Gamma_{k-1}(o)$ and $\alpha\sim\nu$. One should
note that the constants  $\gamma_{k}$ and $\eta_k$ depend only on
strata number.

Then by defining $
\beta_{k}=\gamma_{k}\sqrt{\frac{\sum_{\alpha\in\Gamma_{k}(o)}g_{k,\alpha}^2}
{\sum_{\xi\in\Gamma_{k-1}(o)}g_{k-1,\xi}^2}} $ and
$\alpha_{k}=\eta_k$ for all
$\alpha\in\Gamma_{k}(o),\xi\in\Gamma_{k-1}(o)$ and
$\xi\sim\alpha$, the three-term recursion relations (\ref{tree1})
reduce to those given in  (\ref{trt}).

Therefore similar to the QD case, the adjacency matrix takes a
tridiagonal form in the basis $\ket{\phi_k}$ (orthonormal  basis
associated with strata of GQD graphs), consequently these basis
are identical with the orthonormal basis produced by Lanczos
algorithm.

 \vspace{1cm} \setcounter{section}{0}
 \setcounter{equation}{0}
 \renewcommand{\theequation}{B-\roman{equation}}
  {\Large{Appendix B}}

Here in this appendix we first prove that in GQD graphs, the ratio
of amplitude of a vertex in a given stratum to its coefficient
appearing in (\ref{unitv1}) is constant, i.e.,
$\frac{\phi_{k,\alpha}}{g_{k,\alpha}}$ is independent of
$\alpha\in\Gamma_{k}(o)$. To do so, let us consider the eigenket
$\ket{\phi_{k}}$ given in (\ref{unitv1}), it is straightforward
to see that, the eigenket $\ket{\phi_{k}}$ together with the
following set of states
\begin{equation}
\ket{\phi_{k,l}^\perp}=\frac{1}{\sqrt{\sum_{\nu\in\Gamma_{k}(o)}}\frac{1}{|g_{k,\nu}|^2}}\sum_{\alpha\in\Gamma_{k}(o)}
\frac{\omega^{l\alpha}}{g_{k,\alpha}}\ket{k,\alpha}, \;\;\;\
l=1,2,...,|\Gamma_{k}(o)|-1,
\end{equation}

form a set of orthonormal basis for a complex space formed by
linear span of eigenkets belonging to stratum $k$ where
$\omega=e^{-\frac{2\pi i}{|\Gamma_{k}(o)|}}$.

The above given states are actually orthogonal to all states of
walk space ($V_w$), since the eigenket of other stratum do not
contain any of  $\ket{k,\alpha}, \alpha\in\Gamma_{k}(o)$.
Therefore, $e^{-iAt}\ket{\phi_o}$ is orthogonal to set of
orthogonal vectors $\ket{\phi_{k,l}^\perp}$, for all
$l=1,2,...,|\Gamma_{k}(o)|-1; k=0,1,...,d$ since it is a state
which remains in $V_w$ for all $t$.
 Now, substituting (\ref{unitv1}) in (\ref{v4}) and
  $\braket{\phi_{k,l}^\perp}{e^{-iAt}|\phi_0}=0,
  l=1,2,...,|\Gamma_{k}(o)|-1$, we get the following set of
  equations for amplitudes of vertices belonging to stratum $k$,
  \begin{equation}\label{eq1}
q_{k}(t)=\frac{1}{\sqrt{\sum_{\nu\in\Gamma_{k}(o)}g_{k,\nu}^2}}
\sum_{\alpha\in\Gamma_{k}(o)}g_{k,\alpha}q_{k,\alpha}(t),
 \end{equation}
\begin{equation}\label{eq2}
0=\frac{1}{\sqrt{\sum_{\nu\in\Gamma_{k}(o)}\frac{1}{g_{k,\nu}^2}}}\sum_{\alpha\in\Gamma_{k}(o)}
\frac{\omega^{-l\alpha}}{g_{k,\alpha}} q_{k,\alpha}(t), \;\;\;\;\
l=1,2,...,|\Gamma_{k}(o)|-1,
\end{equation}
where $q_{k,\alpha}(t)$ denotes the amplitude of vertex
$\alpha\in\Gamma_{k}(o)$. To solve equations (\ref{eq1}) and
(\ref{eq2}), first we multiply equations (\ref{eq2}) by
$\omega^{l\nu}$ and sum over $l=1,2,...,|\Gamma_{k}(o)|-1$, where
by using the identity
$\sum_{l=0}^{|\Gamma_{k}(o)|-1}\omega^{l(\nu-\alpha)}=|\Gamma_{k}(o)|\delta_{\alpha\nu}$,
we get for $\nu\neq\alpha$
\begin{equation}
\frac{q_{k,\alpha}(t)}{g_{k,\alpha}}=\frac{1}{|\Gamma_{k}(o)|}\sum_{\nu\neq\alpha}\frac{q_{k,\nu}(t)}{g_{k,\nu}},
\;\;\;\;\ \mbox{for all} \;\ \alpha\in\Gamma_{k}(o).
\end{equation}

Above equations imply that
$\frac{q_{k,\alpha}(t)}{g_{k,\alpha}}=\frac{q_{k,\xi}(t)}{g_{k,\xi}}=B_k$
for all $\alpha,\xi\in\Gamma_{k}(o)$ where, $B_k$ is some constant
independent of vertices of stratum $k$, and it can be determined
by substituting $q_{k,\alpha}(t)=B_k g_{k,\alpha}$ in (\ref{eq1})
as
\begin{equation}
B_k=\frac{1}{\sqrt{\sum_{\nu\in\Gamma_{k}(o)}g_{k,\nu}^2}}q_{k}(t).
\end{equation}
Therefore, probability amplitude of observing walk at given vertex
is proportional to its coefficient $g_{k,\alpha}$ and it can be
written in terms of amplitude of the $k$-th stratum  $q_{k}(t)$ as
\begin{equation}
q_{k,\alpha}(t)=\frac{g_{k,\alpha}}{\sqrt{\sum_{\nu\in\Gamma_{k}(o)}g_{k,\nu}^2}}q_{k}(t).
\end{equation}

In QD graphs we have $g_{k,\alpha}=1$, for all
$\alpha\in\Gamma_{k}(o)$, hence vertices belonging to the same
stratum, have the same amplitude which is in agreement with the
result of appendix I of Ref.\cite{js}.

In non-GQD type graphs, the coefficients of unit vectors
$\ket{\phi_k}$ do not satisfy the conditions (\ref{A1})-(\ref{A3})
,
 and we can not obtain  vectors orthogonal
to $V_{walk}$ by the above explained  prescription of GQD graphs.
Therefore, one should use Lanczos algorithm for obtaining $n$
independent linear equations, where the amplitudes of vertices of
the graph can be deteremined by solving them. Let the Krylov
subspace generated by the adjacency matrix $A$ and starting site
$\ket{\phi_0}$ has dimension $d$, then we will have $d$ unit
vectors of strata produced from Lanczos algorithm applied to the
pair $(A,\ket{\phi_0})$ (one should note that the walk space
$V_{walk}$ is generated by applying the Lanczos algorithm to
adjacency matrix and starting site of the walk ). In the most
cases, the dimension of $V_{walk}$ is less than the number of
vertices ($d<n$), i.e., the Lanczos algorithm applied to the pair
$(A,\ket{\phi_0})$, dose not produce the enough basis, therefore
for obtaining remaining equations we choose  new reference states
orthogonal to walk space $V_{walk}$ and then we apply the Lanczos
algorithm  to the adjacency matrix with new reference states,
respectively. In the following, we explain the procedure in
details for the following example.

 \textbf{Example}

 We consider tree graph of Fig.4, with six vertices and complete orthonormal basis
$$\{\ket{1},\ket{2},\ket{3},\ket{4},\ket{5},\ket{6}\},$$ where
vertex $\ket{1}$  is considered as starting site of the walk. We
apply the Lanczos algorithm to adjacency matrix  $A$ and starting
site $\ket{\phi_0}=\ket{1}$, where orthonormal basis and
coefficients $\alpha_k, \beta_k$ produced from Lanczos algorithm
are
$$
\ket{\phi_0}=\ket{1}, \;\;\;\;\;\
\ket{\phi_1}=\frac{1}{\sqrt{3}}(\ket{2}+\ket{3}+\ket{4})
$$
$$
\ket{\phi_2}=\frac{1}{\sqrt{2}}(\ket{5}+\ket{6}), \;\;\;\;\;\;\
\ket{\phi_3}=\frac{1}{\sqrt{6}}(-2\ket{2}+\ket{3}+\ket{4}),
$$
\begin{equation}
\beta_1=\sqrt{3}, \;\;\;\;\ \beta_2=\sqrt{2/3}, \;\;\;\;\
\beta_3=\sqrt{1/3}; \;\;\;\;\
\alpha_1=\alpha_2=\alpha_3=\alpha_4=0,
\end{equation}
respectively. One can  straightforwardly show that the
corresponding Stieltjes transform of $\mu$ and spectral
distribution are
$$G_{\mu}(z)=\frac{z^3-(1+\sqrt{2})z/\sqrt{3}}{z^4-(4+\sqrt{2})z^2/\sqrt{3}+1},
$$
$$
\mu=0.2851952676(\delta(x-1.662563892)+\delta(x+1.662563892))
$$
\begin{equation}
+0.2148047323(\delta(x-0.6014806445)+\delta(x+0.6014806445),
\end{equation}
respectively, which yield the following probability amplitudes of
walk at $k$-th stratum at time $t$, for $k=0,1,2,3$
$$
q_0(t)=\int_{R}e^{-ixt}\mu(dx)= 0.2851952676\cos(1.662563892t)
+0.2148047323\cos(0.6014806445t).
$$
$$
q_1(t)=\frac{1}{\sqrt{3}}\int_{R}e^{-ixt}P'_1(x)\mu(dx)=\frac{1}{\sqrt{3}}\int_{R}xe^{-ixt}\mu(dx)
$$
$$
=\frac{2}{i\sqrt{3}}[0.2851952676\sin(1.662563892t)+
0.2148047323\sin(0.6014806445t)],
$$
$$
q_2(t)=\frac{1}{\sqrt{2}}\int_{R}e^{-ixt}P'_2(x)\mu(dx)=\frac{1}{\sqrt{2}}\int_{R}(x^2-\sqrt{3})e^{-ixt}\mu(dx)
$$$$
=\frac{1}{\sqrt{2}}[0.2943408772\cos(1.662563892t)
-.2943408762\cos(0.6014806445t)],
$$
$$
q_3(t)=\frac{\sqrt{3}}{\sqrt{2}}\int_{R}e^{-ixt}P'_3(x)\mu(dx)=
\frac{\sqrt{3}}{\sqrt{2}}\int_{R}(x^3-\frac{3+\sqrt{2}}{\sqrt{3}}x)e^{-ixt}\mu(dx)
$$
\begin{equation}\label{tree}
= \frac{6}{i\sqrt{6}}[0.102214289\sin(1.662563892t)-
0.2825324240\sin(0.6014806445t)].
\end{equation}
Obviously, we need two extra  equations for obtaining amplitudes
on sites of the graph. According to the above explained
prescription we can consider
\begin{equation}
\ket{\psi_0}=\frac{1}{\sqrt{2}}(\ket{5}-\ket{6}),
\end{equation}
as new reference state ($\ket{\psi_0}\in V_{walk}^{\perp}$) and
then by applying Lanczos algorithm to the pair ($A, \ket{\psi_0}$)
we obtain
\begin{equation}
\ket{\psi_1}=\frac{1}{\sqrt{2}}(\ket{3}-\ket{4}),
\end{equation}
which leads to two following extra equations
$$
\braket{\psi_0}{e^{-itA}|\phi_0}=0,
$$
\begin{equation}
\braket{\psi_1}{e^{-itA}|\phi_0}=0.
\end{equation}
Now, by solving the above six equations, one can obtain amplitudes
of continuous-time quantum walk on vertices of the graph as
$$\braket{1}{e^{-iAt}\mid\phi_0}=q_0(t),$$
$$\braket{2}{e^{-iAt}\mid\phi_0}=\frac{1}{\sqrt{3}}(q_1(t)-\sqrt{2}q_3(t)),$$
$$\braket{3}{e^{-iAt}\mid\phi_0}=\braket{4}{e^{-iAt}\mid\phi_0}=
\frac{1}{\sqrt{3}}(q_1(t)+\frac{1}{\sqrt{2}}q_3(t)),$$
\begin{equation}
\braket{5}{e^{-iAt}\mid\phi_0}=\braket{6}{e^{-iAt}\mid\phi_0}=\frac{1}{\sqrt{2}}q_2(t),
\end{equation}
where $q_0(t), q_1(t), q_2(t)$ and $q_3(t)$ have been given in
Eq.(\ref{tree}).

\newpage
{\bf Figure Captions}

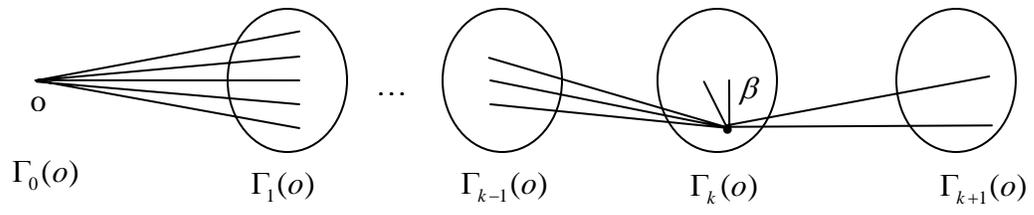
{\bf Figure.1:} Shows the stratification with respect to distance
function.

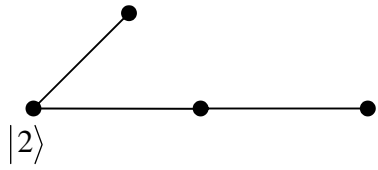
{\bf Figure.2:} Shows the finite path graph of $\textsf{P}_n$,
where walk starts at vertex $2$.

{\bf Figure.3:} Kite graph with $k=2$ and $n=6$, where walk
starts at vertex (0,0). All vertices lying on a given vertical
dashed line belong to the same statum

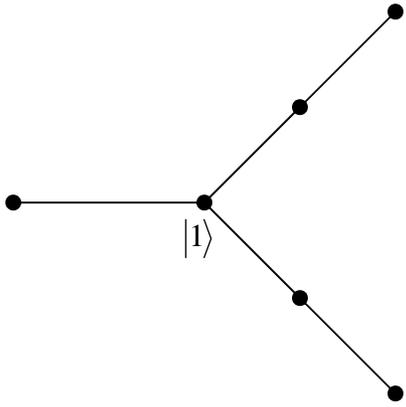
{\bf Figure.4:} Shows the tree graph, where walk starts at vertex
$1$.
\end{document}